\documentstyle[preprint,aps]{revtex}

\tightenlines

\begin{document}
\draft
\title{Strange Mesonic Transition Form Factor\\
in \\
the Chiral Constituent Quark Model}
\author{Hiroshi Ito}
\address{Department of Physics, the George Washington University\\
Washington, DC 20052 USA}
\author{M. J. Ramsey-Musolf\thanks{%
On leave from the Department of Physics, University of Connecticut, Storrs,
CT 06268 USA.}}
\address{Institute for Nuclear Theory, University of Washington, \\
Seattle, WA 98195 USA}
\maketitle

\begin{abstract}
The form factor $g_{\rho \pi }^{(S)}(Q^{2})$ of the strange vector current
transition matrix element $\langle\rho| \bar{s}\gamma_\mu s|\pi\rangle$ is
calculated within the chiral quark model. A strange vector current of the
constituent $U$- and $D$-quarks is induced by kaon radiative corrections and
this mechanism yields the nonvanishing values of $g_{\rho \pi}^{(S)}(0)$.
The numerical result at the photon point is consistent with the one given by
the $\phi $-meson dominance model, but the fall-off in the $Q^{2}$%
-dependence is faster than the monopole form factor. Mesonic radiative
corrections are also examined for the electromagnetic $\rho $-to-$\pi $ and $%
K^{*}$-to-$K$ transition amplitudes.
\end{abstract}

\hfill

\hfill

\hfill

\hfill

\hfill

\hfill

\hfill

\hfill

\hfill

\hfill

\hfill

\hfill

\hfill

\hfill

\hfill

\noindent {INT \#DOE/ER/40561-337-INT97-00-177}\newpage
 INTRODUCTION

The strangeness content of the nucleon is a subject of growing interest in
nuclear and particle physics. From an empirical analysis\cite{A} of the
baryon octet masses, it is inferred that a significant portion of the
nucleon mass is generated by the strange quarks: $m_{S}<N|\overline{s}s|N>$ $%
\simeq 130$ MeV \cite{B}. Nevertheless, conventional nucleon and nuclear
structure models have generally omitted explicit consideration of $s\bar{s}$
degrees of freedom. The validity of this practice will be tested, in part,
by several experiments measuring the distributions of the strange-quark
vector ($\overline{s}\gamma ^{\mu }s$) and axial vector ($\overline{s}\gamma
^{\mu }\gamma ^{5}s$) currents inside the nucleon by means of
parity-violating (PV) elastic electron-proton or electron-nucleus scattering 
\cite{C,C1,D,E,F} and by elastic neutrino-nucleon scattering \cite{G},
respectively. PV elastic electron scattering from ($J^{\pi },T$)=($0^{+},0$)
nuclei is of particular interest because in this case the strangeness
contribution to the helicity asymmetry nominally depends only on the strange
electric form factor of the nucleon $G_{E}^{S}$ \cite{H} (relativistic
corrections introduce a weak dependence on the magnetic form factor, $%
G_{M}^{S}$ \cite{L}). This situation follows from the spin-isospin quantum
numbers for this particular type of nucleus. Moreover, one expects nuclear
structure effects, such as the isospin-mixing in the wave function, to be
negligible in the cases of the light nuclei \cite{I}. With these
considerations in mind, direct measurements of $G_{E}^{S}$ are planned at
TJNAF with $^{4}$He nucleus as the target \cite{D,E}.

Recent theoretical studies, however, have pointed out the prospective
importance of the meson exchange current (MEC) contribution to the $^{4}$He
strangeness form factor. Of particular relevance is the $\rho $-to-$\pi $
transition MEC \cite{J}. Unlike the purely pionic and $N\bar{N}$ MEC's,
which depend on the single nucleon strangeness form factors \cite{L,M}, the $%
\rho$-to-$\pi$ current arises from non-nucleonic strange-quarks through the
amplitude $\langle\rho|\bar{s}\gamma_\mu s|\pi\rangle$. With the use of a
realistic nuclear wave function obtained in the variational Monte Carlo
method\cite{K} , it has been found that the transition MEC and one-body
current may have comparable magnitudes near the expected kinematics of the
experiment of Ref. \cite{D}: $Q^{2}=0.6$ (GeV$/$c)$^{2}$\cite{L}\footnote{%
This result assumes that the single nucleon form factor yields a strangeness
radius of the scale given by pole model calculations \cite{Jaf89}}.
Consequently, a reliable extraction of the nucleon strangeness form factors
from the $^4$He PV asymmetry requires knowledge of both the magnitude of $%
g_{\rho \pi }^{(S)}(Q^{2})$ as well as the degree of model dependence
associated with theoretical estimates.

In this note, we compute $g_{\rho \pi }^{(S)}(Q^{2})$ using a model which
complements those used in previous calculations. Our aim is not to arrive at
a definitive prediction for this form factor, but rather to obtain an
estimate which, when compared to results of other calculations, will help to
determine the scale of model dependence in the corresponding MEC. In
conventional analysis, the MEC operator itself depends on the amplitude of
the $\rho $-to-$\pi $ strange vector current transition: 
\begin{equation}
<\rho (p+q,\varepsilon )|\overline{s}\gamma ^{\mu }s|\pi (p)>=\frac{g_{\rho
\pi }^{(S)}(Q^{2})}{m_{\rho }}\epsilon ^{\mu \nu \alpha \beta }\text{ }%
p_{\nu }\text{ }q_{\alpha }\varepsilon _{\beta }^{*}\text{ }
\end{equation}
\noindent where $\varepsilon _{\beta }$ is the polarization of $\rho $
meson. This amplitude is off-diagonal, so that no conservation principle
constrains the form factor $g_{\rho \pi }^{(S)}(Q^{2})$ at the photon point (%
$Q^{2}=0$). In the absence of its direct measurement one must rely on
theoretical models for $g_{\rho \pi }^{(S)}(Q^{2})$. Ideally, such models
should also yield a reasonable explanation of known observables, such as the
electromagnetic form factor $g_{\rho \pi }^{(EM)}(Q^{2})$. Along these
lines, two theoretical estimates of the strange vector $\rho $-to-$\pi $
transition amplitude have been reported, both using a hadronic framework:
one with the $\phi $-meson dominance model \cite{M} and the other with the $%
K^{*}K$ loop model \cite{N}.

In the present paper, we calculate $g_{\rho \pi }^{(S)}(Q^{2})$ with the
chiral constituent quark model \cite{O}, where the Goldstone bosons couple
to the constituent quarks in a chiral invariant manner. The mass spectrum 
\cite{P} and electromagnetic form factors \cite{Q,R} of nonstrange baryons
and mesons are well explained with the constituent quark model, where the
hadrons are composed of the weekly interacting nonstrange constituent
quarks. With this successful effective description of QCD on the one hand
and the empirical observation \cite{B} of the nucleon's strangeness content
on the other, one would expect the nonstrange constituent quarks may have
complex structure including $\overline{s}s$ pairs as a component \cite{Kap89}%
. In the chiral quark model, this effect is taken into account via kaon
radiative corrections which induce a ``strangeness polarization'' of the
nonstrange quarks: $U\rightarrow K^{+}+S\rightarrow U$ and $D\rightarrow
K^{0}+S\rightarrow D$. This mechanism of flavor changing mesonic radiative
corrections also modifies the electroweak couplings of constituent quarks,
and we examine such effects on both vector strange and electromagnetic
current processes in the same framework. Although the model faces some
conceptual ambiguities \cite{Isg92}, it has, nevertheless, enjoyed
considerable success in describing a variety of low-energy nucleon
properties \cite{O,XQM1,XQM2,XQM3}. We therefore treat it as a useful,
though not definitive, tool for estimating the scale of flavor mixing in
constituent quarks.

In the framework of constituent quark models, the strangeness distribution
in a nonstrange quark can be described in terms the strange vector current
of the $U$ and $D$ quarks, defined as \cite{S1} 
\begin{equation}
J_{S}^{\mu }=f_{1}^{(S)}(Q^{2})\gamma ^{\mu }+\frac{i}{2M}%
f_{2}^{(S)}(Q^{2})\sigma ^{\mu \nu }q_{\nu },
\end{equation}

\noindent where $Q^{2}\equiv {\bf q}^{2}-q_{0}^{2}$ is the momentum
transfer, and $M=M_{u}\simeq M_{d}\sim 300$ MeV is the constituent quark
mass. Since the net strangeness is zero for these quarks, the Dirac form
factor vanishes at the photon point ( $f_{1}^{(S)}(0)=0)$, but the anomalous
magnetic moment is subject to the renormalization ( $f_{2}^{(S)}(0)\neq 0)$
due to the cloud of $K$-mesons.

In the absence of the Dirac form factor, the $\rho $-to-$\pi $ transition
amplitude given by the quark model matrix element of $J_{S}^{\mu }$ is
proportional to $f_{2}^{(S)}(0)$ (up to corrections) at the photon point

\begin{equation}
g_{\rho \pi }^{(S)}(0)=\alpha f_{2}^{(S)}(0)+\cdot \cdot \cdot ,
\end{equation}
\noindent where $\alpha $ depends on the quark distributions in the hadrons.
Higher order terms may arise from many-body currents and correlations. For
purposes of arriving at an estimate, however, retention of only the one-body
terms is sufficient. With the use of the strangeness current in Eq. (2) we
calculate the form factor $g_{\rho \pi }^{(S)}(Q^{2})$ by using the
relativistic quark model wave function. The first step is to evaluate the
constituent quark form factors $f_{i}^{S}(Q^{2})$ by calculating the vertex
correction at the quark coupling to the electroweak neutral boson in Figures
1b-1g (also Fig.1 in \cite{Z,S2}).We subsequently evaluate the matrix
element of $J_{S}^{\mu }$ using appropriate flavor-spin-space wave
functions. A similar approach has been followed in estimating the nucleon's
strange magnetic moment ( $G_{M}^{S}(0)$ ) 
\begin{equation}
G_{M}^{S}(0)=\beta f_{2}^{(S)}(0)+\cdot \cdot \cdot ,
\end{equation}
\noindent where $\beta $ can be obtained from the quark model wave function
of the nucleon \cite{S1,Z,S2}.

\noindent MODEL

The $SU(3)_{L}\times SU(3)_{R}$ chiral symmetry of three-flavor QCD is
spontaneously broken at the level of the quark and gluon degrees of freedom,
yielding the so-called ``soft'' mass ($\sim M$) \cite{O,Politz} and the
appearance of the octet Goldstone bosons. The chiral quark model \cite{O}
incorporates this aspect by introducing the non-linear fields of interacting
Goldstone boson octet ($\varphi $) that couple to the constituent quark
field ($\psi $) 
\begin{equation}
{\cal L}=i\overline{\psi }({\cal D}_{\mu }+V_{\mu })\gamma ^{\mu }\psi -%
\overline{\psi }{\bf M}\psi -g_{A}\overline{\psi }A_{\mu }\gamma ^{\mu
}\gamma ^{5}\psi +\frac{f_{\pi }}{4}tr\{\partial ^{\mu }\Sigma ^{\dagger
}\partial _{\mu }\Sigma \}+\cdot \cdot \cdot
\end{equation}
\noindent where $V_{\mu }=\frac{1}{2}(\xi ^{\dagger }\partial _{\mu }\xi
+\xi \partial _{\mu }\xi ^{\dagger })$, $A_{\mu }=\frac{i}{2}(\xi ^{\dagger
}\partial _{\mu }\xi -\xi \partial _{\mu }\xi ^{\dagger })$ and $\Sigma =\xi
^{2}$ with $\xi =e^{i\varphi /f_{\pi }\text{ }}$. Here, ${\bf M}%
=\{M_{U},M_{D},M_{S}\}$ is the diagonal mass matrix of the constituent
quarks, $f_{\pi }=93MeV$ is the pion weak decay constant, and the covariant
derivative ${\cal D}_{\mu }$ includes the gluon fields. The octet of the
Goldstone boson fields is expressed in a matrix form in the quark-flavor
space, 
\begin{equation}
\varphi =\frac{1}{\sqrt{2}}\left[ 
\begin{array}{ccc}
\frac{1}{\sqrt{2}}\pi ^{0}+\frac{1}{\sqrt{6}}\eta & \pi ^{+} & K^{+} \\ 
\pi ^{-} & \frac{-1}{\sqrt{2}}\pi ^{0}+\frac{1}{\sqrt{6}}\eta & K^{0} \\ 
K^{-} & \overline{K}^{0} & \frac{-2}{\sqrt{6}}\eta
\end{array}
\right] .
\end{equation}
\noindent The strange vector current is obtained by the minimal substitution
of the external source coupling to strangeness, ${\cal Z}_{\mu }$: $\partial
_{\mu }\rightarrow \partial _{\mu }+i\omega _{S}{\cal Z}_{\mu }$ for each
field with the strangeness $\omega _{S}$. The strangeness form factors $%
f_{1}^{(S)}(Q^{2})$ and $f_{2}^{(S)}(Q^{2})$ are calculated from the
diagrams of the $K$-meson loop expansion\cite{S2}, which are the lowest
order in $\frac{1}{N_{C}}$ ($\sim $ $\frac{1}{f_{\pi }^{2}}$ ). The matrix
element of the strange vector current for the $\rho $-to-$\pi $ transition
is given by the sum of Fig.1b,c,d,e,f and Fig.1g, where one of the
nonstrange quarks turns into the strange quark accompanied with the $K$
meson cloud; $U$($D$)$\rightarrow S+K\rightarrow U$($D$), and the external
neutral current couples to the strangeness.

The loops of Figs. 1b-k are U.V. divergent, and one may remove such
divergences through the appropriate higher-order terms in the chiral
Lagrangian (see, {\em e.g.}, Eq. (46) of Ref. \cite{Z}). In general, the
finite components of these counter-terms may be determined from existing
data using chiral symmetry. In the case of the strangeness form factors of a
non-strange constituent quark, however, such a determination is not possible 
\cite{Z}. An alternate strategy is to cut the loop-momentum off at the scale
of the chiral symmetry breaking $\Lambda =\Lambda _{\chi }\sim 1$ GeV. We
implement this procedure in a gauge-invariant way by introducing a monopole
form factor $(\Lambda _{{}}^{2}-\mu _{{}}^{2})/(\Lambda _{{}}^{2}-k^{2})$ at
the quark-meson vertex, where $k$ is the four momentum of the meson and $\mu 
$ is the mass. The gauge invariance is minimally satisfied for the sum of
the amplitudes Fig.1b, c, d and Fig.1e, where the last two terms are
associated with the momentum dependence of the meson-quark vertex. The
quarks also couple to the vector fields of the Goldstone bosons by $%
\overline{\psi }V_{\mu }\gamma ^{\mu }\psi $, and this vertex induces the
amplitudes Fig.1f and Fig.1g in the same order of the expansion. The sum of
these two are gauge invariant by itself.

The hadronic matrix element [Eq. (1)] of the strange vector current in Eq.
(2) is dominated by the anomalous magnetic moment term, and this term is of
the relativistic order ($\sim q/M$). The use of the light-cone quark model
has been quite successful in predicting the electromagnetic\cite{Q,R} and
strangeness \cite{S1,Z} form factors of hadrons including the off-diagonal
meson form factors of the $\gamma +\pi \rightarrow \rho $ and $\gamma
+K\rightarrow K^{*}$ transitions \cite{U,V,V1}, and we follow the method of
Ref\cite{T} in this work. The $+$ component of a vector current ( $%
j^{+}\equiv j^{0}+j^{3}$ ) is commonly used in the light-cone approach, and
with this choice the phenomenological form of the transition amplitude in
Eq. (1) becomes

\begin{equation}
<\rho ^{a}(p^{\prime },\varepsilon )|j^{+}|\pi ^{b}(p)>=\frac{g_{\rho \pi
}^{(S)}(Q^{2})}{m_{\rho }}p^{+}[\varepsilon ^{*}\times {\bf q}]_{z},
\end{equation}

\noindent where $p^{+}\equiv p^{0}+p^{3}$. The four momentum of the $\pi $
meson moving toward the $z$-direction is $p^{\mu }=(m_{\pi }^{2}/p^{+},$ $%
{\bf 0},$ $p^{+})$ and the one of the $\rho $ meson in the final state is $%
p^{\prime \mu }=([{\bf q}_{\bot }^{2}+m_{\rho }^{2}]/p^{+},$ ${\bf q}_{\bot
},$ $p^{+})$ with the light-cone representation of a four-vector $V^{\mu
}=(V_{-},{\bf V}_{\bot },V_{+})$. The photon momentum satisfies $q^{2}=-{\bf %
q}_{\bot }^{2}$ with ${\bf q}_{\bot }=(q_{x},q_{y,}0).$ The transverse ($%
{\bf \varepsilon }_{T}$) and longitudinal (${\bf \varepsilon }_{L}$)
polarization vectors of the $\rho $ meson are

\begin{mathletters}
\label{allequations}
\begin{equation}
\varepsilon _{T}(p^{\prime })=(2{\bf \varepsilon }_{\bot }\cdot {\bf q}%
_{\bot }/p^{+},{\bf \varepsilon }_{\bot },0)
\end{equation}
\begin{equation}
\varepsilon _{L}(p^{\prime })m_{\rho }=([{\bf q}_{\bot }^{2}-m_{\rho
}^{2}]/p^{+},{\bf q}_{\bot },p^{+})
\end{equation}
\noindent where ${\bf \varepsilon }_{\bot }\equiv {\bf \varepsilon }_{\bot
}^{\oplus }=(\frac{-1}{\sqrt{2}},-\frac{i}{\sqrt{2}},0)$ for the positive
helicity state$,$ where ${\bf \varepsilon }_{T}\cdot p^{\prime }={\bf %
\varepsilon }_{L}\cdot p^{\prime }=0$ is satisfied \cite{W}. Note that only
the transversely polarized $\rho $ meson can contribute to the matrix
element (${\bf \varepsilon }_{L}\times {\bf q}_{\bot }=0$). In this frame, a
vector current expressed with the Dirac matrices can be reduced to the Pauli
matrix form$:$ $J_{S}^{+}\Rightarrow j^{+}=2\{f_{1}^{(S)}(Q^{2})I+\frac{i}{2M%
}f_{2}^{(S)}(Q^{2})[{\bf \sigma \times q}_{\bot }]_{z}\}$, and the matrix
element of the transition amplitude is explicitly calculated with the use of
the quark model wave functions ($\Psi _{\pi }$ and $\Psi _{\rho }$)

\end{mathletters}
\begin{equation}
\begin{array}{c}
\noindent <\rho ^{a}(p^{\prime },\varepsilon )|j^{+}|\pi ^{b}(p)>=\frac{%
2p^{+}}{16\pi ^{3}}\int dx\int d^{2}k_{\bot }\frac{1}{x(1-x)} \\ 
\\ 
\times \sum\limits_{s_{a},s_{a^{\prime }},s_{b}}\Psi _{\rho }^{\dagger
}(s_{b},s_{a^{\prime }})\{f_{1}^{(S)}(Q^{2})I+\frac{i}{2M}f_{2}^{(S)}(Q^{2})[%
{\bf \sigma \times q}_{\bot }]_{z}\}_{s_{a^{\prime }}s_{a}}\Psi _{\pi
}(s_{a},s_{b})+(a\leftrightarrow b).
\end{array}
\end{equation}

\noindent Here, momenta of quark (a) and antiquark (b) in the initial state
are defined by ${\bf p}_{a}=x_{a}{\bf p}{\bf +}{\bf k}_{a\bot }$ and ${\bf p}%
_{b}=x_{b}{\bf p+k}_{b\bot },$ where $x=x_{a}$ and $x_{b}=1-x_{a}$ are the
fractions of their longitudinal momenta and the transverse momenta are given
by ${\bf k}_{a\bot }=-{\bf k}_{b\bot }=$ ${\bf k}_{\bot }-\frac{1}{2}%
(1-x_{a}){\bf q}_{\bot }$ . With a convenient choice of the photon momentum $%
{\bf q}_{\bot }=(Q,0,0)$, the wave function of the transversely polarized $%
\rho $ meson $\Psi _{\rho }(s_{b},s_{a^{\prime }})$ is proportional to the $%
y $-component of the polarization vector,

\begin{equation}
\Psi _\rho ^{\dagger }(s_b,s_{a^{\prime }})={\bf \varepsilon }_{\bot y}^{*}%
{\cal N}_\rho <s_b|a_0I+a_1\sigma _1+a_2\sigma _2+a_3\sigma _3|s_{a^{\prime
}}>\chi _{cf}^{*}(\rho )\varphi _\rho (M_0^{\prime 2})
\end{equation}

\noindent where the quark-spin configuration of the vector meson, $%
<s_{b}|\cdot \cdot \cdot |s_{a^{\prime }}>,$ is obtained from the Melosh
transformation \cite{X} of the spinor matrix element [$\overline{u}\gamma
^{\mu }v$]\cite{T}. The coefficients $a_{n}$ are given by $%
a_{0}=i(k_{x}^{\prime 2}-k_{y}^{\prime 2}-e_{1}^{\prime }e_{2}^{\prime
})/D^{\prime },$ $a_{1}=(e_{1}^{\prime }-e_{2}^{\prime })k_{y}^{\prime
}/D^{\prime }$, $a_{2}=-(e_{1}^{\prime }+e_{2}^{\prime })k_{x}^{\prime
}/D^{\prime }$ and $a_{3}=-2k_{x}^{\prime }k_{y}^{\prime }/D^{\prime }$ with 
$D^{\prime }=\sqrt{2}d_{1}^{\prime }d_{2}^{\prime },$ $d_{i}^{\prime }=\sqrt{%
k_{\bot i}^{\prime 2}+e_{i}^{\prime 2}}$ and $e_{i}^{\prime
}=m_{i}+x_{i}M_{0}^{\prime }.$ Here, $M_{0}^{^{\prime
}2}=\sum\limits_{i=a}^{b}(k_{i\bot }^{\prime 2}+m_{i}^{2})/x_{i}$ is the
invariant mass of the quark-antiquark system. The momentum distribution is
denoted by $\varphi _{\rho }(M_{0}^{^{\prime }2})$, and we use the gaussian
function $\varphi _{\rho }(x)=\exp (-\frac{x^2}{2\gamma ^{2}})$ \cite{Y}.
The color and flavor wave function is denoted by $\chi _{cf}(\rho ),$ for
example $\chi _{cf}(\rho ^{+})=\chi _{cf}(\pi ^{+})=\frac{1}{\sqrt{N_{C}}}%
\left[ 
\begin{array}{ll}
0 & 1 \\ 
0 & 0
\end{array}
\right] $for the charged $\rho $ and $\pi $ mesons, and the normalization
factor ${\cal N}_{\rho }$ is determined by the charge normalization for the
charged $\rho $ meson. The wave function of the pion is

\begin{equation}
\Psi _\pi (s_a,s_b)={\cal N}_\pi <s_a|b_0I+b_1\sigma _1+b_2\sigma
_2+b_3\sigma _3|s_b>\chi _{cf}(\pi )\varphi _\pi (M_0^2)
\end{equation}

\noindent where the coefficients $b_{n}$ are $b_{0}=-i(e_{1}+e_{2})k_{x}/D,$ 
$b_{1}=0,$ $b_{2}=(k_{\bot }^{2}-e_{1}e_{2})/D$ and $%
b_{3}=-(e_{1}+e_{2})k_{y}/D$ . In the present work based on the chiral quark
model Lagrangian, the pion is considered to be point-like object, with
pseudovector coupling to the quark-antiquark pair: $\frac{g_{A}}{f_{\pi }}
p_{\mu }\gamma ^{\mu}\gamma ^{5}$. Consistency between these properties and
the formalism of Eq. (9) requires that one set $\varphi _{\pi
}(M_{0}^{2})=const$ and ${\cal N}_{\pi }=\frac{g_{A}}{f_{\pi }}\sqrt{N_{c}}$
(the pion wave function is defined with the explicit color dependence $\chi
_{cf}(\pi )\sim 1/\sqrt{N_{c}}$).

\noindent RESULTS AND DISCUSSION

We start with a discussion of the electromagnetic radiative decay of a
vector to pseudoscalar meson. The numerical results for the $\rho
^{+}\rightarrow \pi ^{+}+\gamma ,$ $K^{0*}\rightarrow K^{0}+\gamma $ and $%
K^{+*}\rightarrow K^{+}$ +$\gamma $ processes are presented in Table I,
where we observe the major contribution from the simple quark model matrix
element Fig.1a obtained with the electromagnetic current operator of
point-like quark $J^{\mu }=e_{q}\gamma ^{\mu }$. To obtain an idea of the
sensitivity of these results to various model parameters, we vary the quark
mass by 10\% (from $M=$330 MeV to 300 MeV). The results for the amplitudes
change by about $5-8\%$. Similarly, varying the harmonic oscillator
parameter of the wave function from $\gamma =0.64GeV$ to $0.70GeV$ reduces
the amplitudes about $3-5\%.$ We note that there remains an ambiguity of
sign for the transition amplitudes, which depends on the relative phases of
the wave function normalizations; $g_{\rho \pi }^{(EM)}\sim {\cal N}_{\rho }%
{\cal N}_{\pi }$ $.$ We cannot fix this phase from the experimental data for
the decay rates $\Gamma _{\exp }\sim |g_{\rho \pi }^{(EM)}|^{2},$ and we
chose ${\cal N}_{\rho }$ and ${\cal N}_{\pi }$ to be real and positive in
this work. The same convention is used for the calculation of the
strangeness form factor $g_{\rho \pi }^{(S)}(Q^{2}).$

The lowest order ${\cal O}(1/\Lambda _{\chi }^{2})$ mesonic radiative
corrections to the matrix element of Fig. 1a are calculated from the ten
diagrams Fig.1b-Fig.1k. The pseudoscalar meson loops include the entire
octet of the Goldstone bosons. These loops generate about a $6-10\%$
correction to the contribution from Fig. 1a, as indicated in the third
column of Table I. This correction is similar in magnitude to the variations
induced by $10\%$ changes of model parameters. (The predicted decay width $%
\Gamma _{th\text{ }}$ given in the fourth column of Table I takes into
account the mesonic radiative corrections.)

For the case of the strange vector current $\rho $-to-$\pi $ transition, we
evaluate the six diagrams of Fig.1b-Fig.1g, where the loops contain $K$
mesons only. The other diagrams (1a, h-k) do not contribute. The strangeness
form factors of the $U$- and $D$-quarks are calculated with the cut-off mass
of $\Lambda =1.0$ GeV (1.2 GeV). The Dirac form factor $f_{1}^{(S)}(Q^{2})$
vanishes at $Q^{2}{=}0$ and the strange anomalous moment is $%
f_{2}^{(S)}(0)=-0.030$ ($-0.041$)\cite{S2}. The momentum dependence of the
strange vector $\rho $ -to- $\pi $ transition $g_{\rho \pi }^{(S)}(Q^{2})$
is plotted in Fig.2, where we actually present the scaled values $\overline{g%
}_{\rho \pi }^{(S)}(Q^{2})=$ $g_{\rho \pi }^{(S)}(Q^{2})/g_{\rho \pi
}^{(S)}(0)$. The value at the photon point is $g_{\rho \pi
}^{(S)}(0)=-0.21(-0.27)$ for $\Lambda =1.0$ GeV $(1.2$ GeV). Here the solid
line is the prediction of the present work with $\Lambda =1.0$ GeV. In the
vector ($\phi $) meson dominance model (VMD), this form factor can be
parametrized in a monopole function 
\begin{equation}
g_{\rho \pi }^{(S)}(Q^{2})=g_{\rho \pi }^{(S)}(0)\frac{m_{\phi }^{2}}{%
Q^{2}+m_{\phi }^{2}}.
\end{equation}
\noindent With the empirical values of the hadronic ($\phi \rightarrow \rho
\pi $ )and leptonic ( $\phi \rightarrow l^{+}l^{-}$) decay rates the VMD
model yields the magnitude $|g_{\rho \pi }^{(S)}(0)|=0.21$ \cite{N}, and
this value has been used in calculating the $\rho \pi $-strange MEC
contribution to the PV $^{4}$He asymmetry \cite{L,M}. The VMD form factor
Eq. (12) is also shown in the Fig.2 (dotted dash line) for the comparison.
Because of the composite nature of the $\rho $-meson, the fall-off in the
quark model form factor is faster that the one in the VMD model. At the
kinematics of experiment\cite{D} ($Q^{2}=0.6$ GeV$^{2}/c^{2}$), the chiral
quark model prediction is about $30\%$ smaller than the one in the VMD model.

To summarize: We have calculated the vector strangeness form factor $g_{\rho
\pi }^{(S)}(Q^{2})$ and electromagnetic decay amplitudes for $\rho \to \pi
\gamma $ and $K^{*}\to K\gamma $ within the chiral constituent quark model.
The mesonic radiative corrections to the constituent quark currents --
generated by the octet of the Goldstone bosons -- are consistently taken
into account for both processes. In the case of the EM transitions, the
corrections are roughly $\leq 10\%$ of the leading term (Fig. 1a), as one
would expect on general grounds \cite{O}. The resultant predictions are in
good agreement with experimental values. On the other hand, the process of
Fig. 1a does not contribute in the strangeness case, and the leading order
contribution arises from the kaon radiative corrections: $U(D)\rightarrow
K+S\rightarrow U(D)$. The nonstrange quarks acquire the strange anomalous
moment $f_{2}^{(S)}(0)$, and we are able to relate the strangeness form
factor, $g_{\rho \pi }^{(S)}(0)\propto $ $f_{2}^{(S)}(0)$ in the present
approach. Our result for $g_{\rho \pi }^{(S)}(Q^{2})$ differs by roughly $%
30\%$ from other model estimates at the kinematics of the future $^{4}$He PV
experiment \cite{D}. We expect the theoretical uncertainty in the
corresponding MEC correction to be at least as large as the degree of
model-dependence in $g_{\rho \pi }^{(S)}(Q^{2})$.

\subsection{ACKNOWLEDGMENTS}

We would like to thank N. Isgur for useful discussions. H.I wishes to thank
for the hospitality to R. Seki and the other members of W. K. Kellogg
Radiation Laboratory at Caltech, where the most of his work was completed.
M.J.R-M has been supported in part under U.S. Department of Energy Contract
No. DE-FG06-90ER40561 and through the National Science Foundation National
Young Investigator Program.

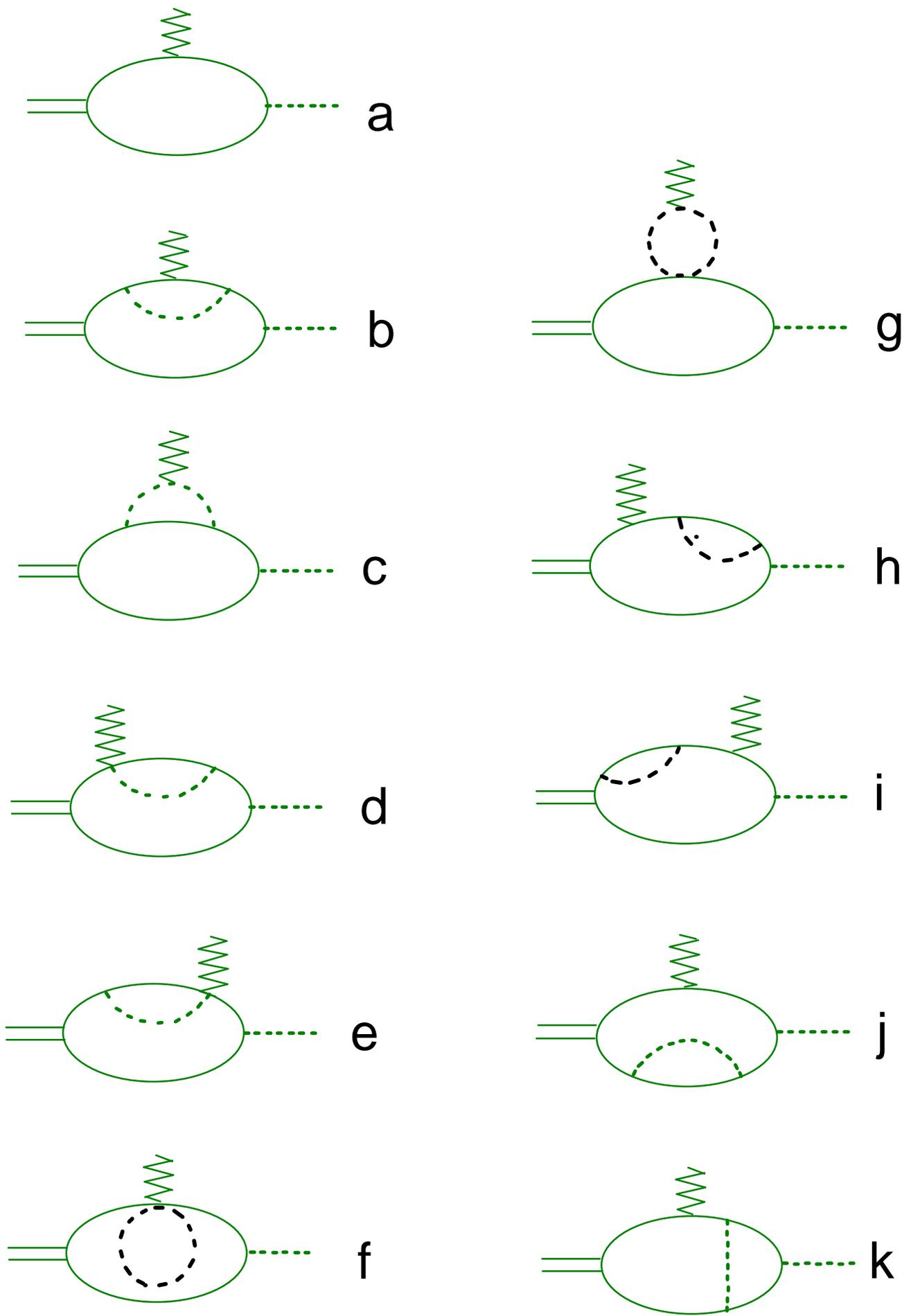
\begin{figure}[tbp]
\caption{Quark loop diagrams for the transition form factor of the vector
(double line)-to- pseudoscalar meson (dot line), where the wavy line is the
electroweak neutral boson or photon. The solid line is the constituent
quark. }
\label{fig1}
\end{figure}

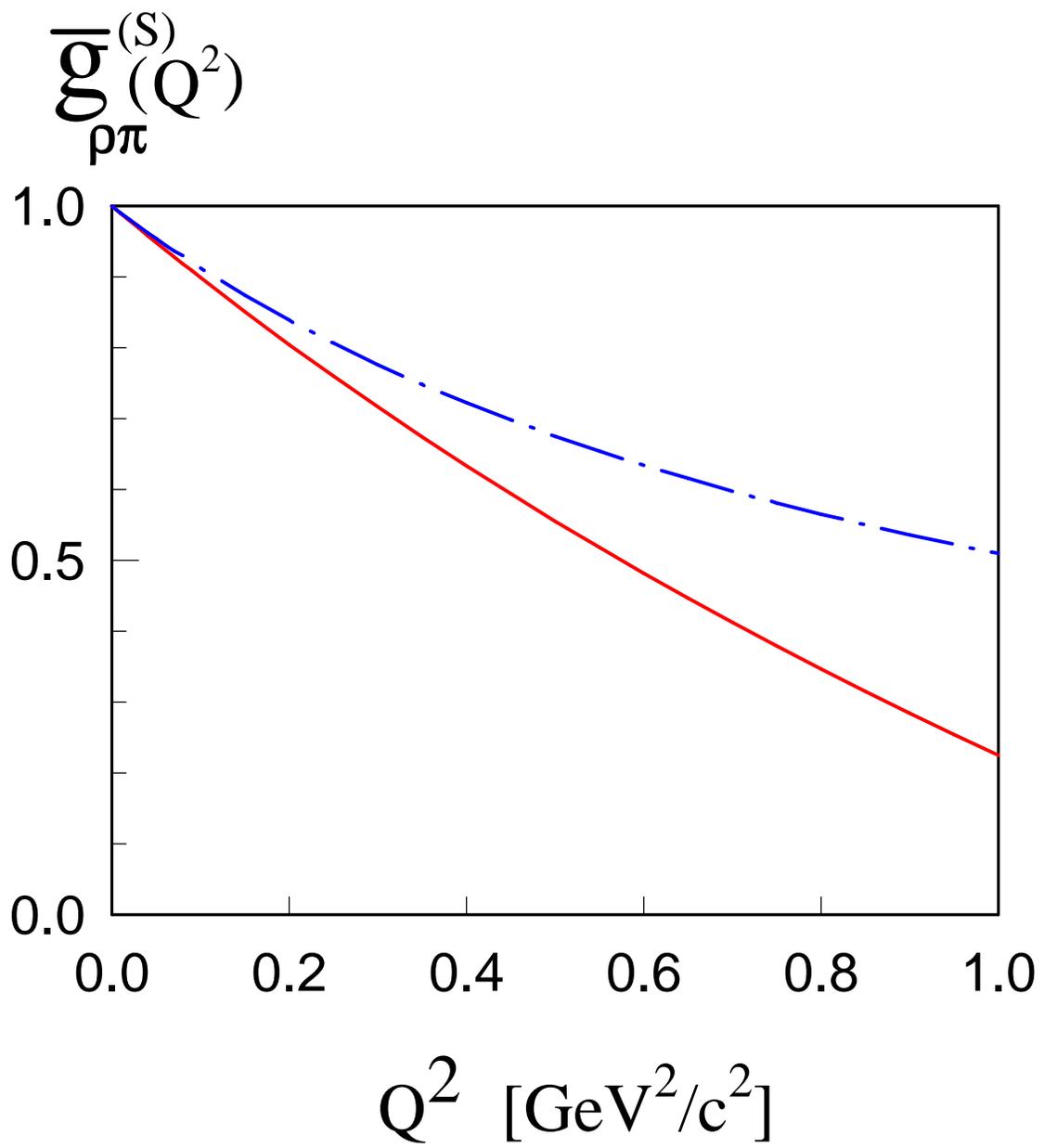
\begin{figure}[tbp]
\caption{Form factor of the strange vector current $\rho$-to-$\pi$
transition defined by $\overline{g}^{(S)}_{\rho\pi}(Q^2)$= $%
g^{(S)}_{\rho\pi}(Q^2) / g^{(S)}_{\rho\pi}(0)$ . The solid line is the
present result with $\Lambda=1GeV$. The monopole form facor of the $\phi$
meson dominance model of Ref. \protect\cite{N} (dotted-dash line) is shown
for comparison. }
\label{fig2}
\end{figure}
\begin{table}[tbp]
\caption{Electromagnetic radiative transition amplitude of the vector ($V$)
-to- pseudoscalar ($P$) meson $g^{(th)}_{VP}$ ( $g^{(0)}_{VP}$ ) with
(without) the mesonic radiative corrections. $m_{V}$ is the mass of the
vector meson. The theoretical ($\Gamma_{th}$) and experimental ($%
\Gamma_{exp} $) decay widths are shown. The harmonic ocilator parameter is $%
\gamma=0.64$ GeV and the quark masses are $M_S=0.45$ GeV and $M_U=M_D=0.33$
GeV. }
\label{table1}
\begin{tabular}{ccccc}
$V \rightarrow P+\gamma $ & $\frac {g^{(0)}_{VP}}{m_V}[GeV^{-1}]$ & $\frac {%
g^{(th)}_{VP}}{m_V}[GeV^{-1}]$ & $\Gamma_{th}[MeV]$ & $\Gamma_{exp}[MeV]$ \\ 
\tableline $\rho^{+} \rightarrow \pi^{+}+\gamma $ & 0.71 & 0.75 & 0.069 & 
0.068 $\pm $ 0.007 \\ 
$K^{+*} \rightarrow K^{+}+\gamma $ & 0.80 & 0.89 & 0.057 & 0.050 $\pm $ 0.005
\\ 
$K^{0*} \rightarrow K^{0}+\gamma $ & -1.20 & -1.30 & 0.122 & 0.117 $\pm $
0.01
\end{tabular}
\end{table}

\end{document}